\documentclass[12pt,a4paper]{article}
\begin{document}
\begin{titlepage}
\begin{flushright}
LNF--00/031(P)\\
UAB--FT--498\\
hep-ph/0012049\\
December 2000
\end{flushright}
\vspace*{1.6cm}

\begin{center}
{\Large\bf Radiative $VP\gamma$ transitions and $\eta$-$\eta^\prime$ mixing}\\
\vspace*{0.8cm}

A.~Bramon$^1$, R.~Escribano$^2$ and M.D.~Scadron$^{3,4}$\\
\vspace*{0.2cm}

{\footnotesize\it 
$^1$Departament de F\'{\i}sica, Universitat Aut\`onoma de Barcelona, 
E-08193 Bellaterra (Barcelona), Spain\\
$^2$INFN-Laboratori Nazionali di Frascati, P.O.~Box 13, I-00044 Frascati, Italy\\
$^3$Physics Department, University of Arizona, Tucson, AZ 85721, USA\\
$^4$IFAE, Universitat Aut\`onoma de Barcelona, E-08193 Bellaterra (Barcelona), Spain}
\end{center}
\vspace*{1.0cm}

\begin{abstract}
A value for the $\eta$-$\eta^\prime$ mixing angle is extracted from the data on 
$VP\gamma$ transitions using simple quark-model ideas.
The set of data covers {\it all} possible radiative transitions between the 
pseudoscalar and vector meson nonets.
Two main ingredients of the model are the introduction of flavour-dependent overlaps 
for the various $q\bar{q}$ wave functions and the use of the quark-flavour basis to 
describe the $\eta$-$\eta^\prime$ system. 
In this basis the mixing angle is found to be $\phi_P=(37.7\pm 2.4)^\circ$.
\end{abstract}
\end{titlepage}

\section{Introduction}
\label{intro}
Radiative transitions between pseudoscalar ($P$) and vector ($V$) mesons have been 
a classical subject in low-energy hadron physics for more than three decades. 
The study of the $VP\gamma$ couplings, governed by the magnetic dipole ($M1$) emission 
of a photon, played a major r\^ole when the basis of the quark model and 
$SU(3)$-symmetry were established, as well as when trying to understand their 
symmetry-breaking mechanisms. 
The pioneering work by Becchi and Morpurgo \cite{BM} successfully explained the 
specific $\omega\rightarrow\pi^0\gamma$ decay rate in terms of the $u,d$ 
quark magnetic moments $\mu_{u,d}$, as deduced from the measurable magnetic moments 
of the nucleons $\mu_{p,n}$.  
Extension to other $V\rightarrow P\gamma$ and $P\rightarrow V\gamma$ radiative decays 
was soon performed exploiting $SU(3)$-symmetry relations as described, for instance, 
in the concise comment by Isgur \cite{Isg} or in the review by O'Donnell \cite{OD} 
where symmetry-breaking effects are also discussed. 
Among the latter, those related to the $\eta$-$\eta^\prime$ system turn out to be 
particularly interesting and have recently heightened the theoretical activity on the 
$VP\gamma$ magnetic dipole transitions \cite{BGP}--\cite{Fel}.

From the experimental point of view, the Novosibirsk CMD-2 \cite{CMD-2} and 
SND \cite{SND} Collaborations have reported very recently accurate and consistent 
results on the various $V\rightarrow P\gamma$ radiative decays and, in particular, 
on the poorly known $\phi\rightarrow\eta^\prime\gamma$ branching ratio.
This latter $e^+e^-$-annihilation result complements older data on the other two 
$\eta^\prime$-meson radiative transitions, $\eta^\prime\rightarrow\rho\gamma,
\omega\gamma$, previously measured through $\gamma\gamma$-interactions.
For the first time we have a well established and consistent set of data covering the 
$V\rightarrow P\gamma$ and $P\rightarrow V\gamma$ radiative decays with 
$V=\rho^0, \omega, \phi$ and $P=\pi^0, \eta, \eta^\prime$, 
as shown by the current PDG edition \cite{PDG00}. 
This set of data is completed by the $\rho^+\rightarrow\pi^+\gamma$,  
$K^{*+}\rightarrow K^+\gamma$ and $K^{*0}\rightarrow K^0\gamma$ 
transitions measured by the Primakoff-effect and thus affected by larger uncertainties. 
Globally, these experimental results represent an exhaustive and useful set of data 
covering {\it all} the twelve possible radiative transitions between the pseudoscalar 
and the vector meson nonets \cite{PDG00}.
Moreover, the Frascati $\phi$-factory DA$\Phi$NE \cite{daphne95:juliet} is expected to 
improve this situation quite soon.

Six of the just mentioned radiative transitions involve $\eta$ or $\eta^\prime$ mesons 
and contain valuable information on the properties of the $\eta$-$\eta^\prime$ system 
and their mixing pattern. 
Interest on this issue has been recently renewed by Leutwyler {\it et al.}~\cite{Leu}
in their analysis of the $f_\eta$ and $f_{\eta^\prime}$ decay constants governing the 
$\eta, \eta^\prime\rightarrow\gamma\gamma$ transitions. 
In their effective chiral Lagrangian context, two mixing angles are required to express 
$f_\eta$ and $f_{\eta^\prime}$ in terms of the octet and singlet decay constants 
$f_8$ and $f_0$, the canonical treatment with one single mixing angle being recovered 
only in the good $SU(3)$ limit. 
Feldmann and collaborators \cite{FKS,Fel} have confirmed that a two mixing angle analysis 
is more coherent than the conventional one, but, on the other side, they have also 
reopened the possibility of a single angle description provided one abandons the 
octet-singlet basis and works in the quark-flavour basis. 
In this case one has 
\begin{equation}
\label{mixingP}
\begin{array}{l}
|\eta\rangle\equiv\cos\phi_P|\eta_{NS}\rangle -\sin\phi_P|\eta_S\rangle\ ,\\[2ex]
|\eta^\prime\rangle\equiv\sin\phi_P|\eta_{NS}\rangle +\cos\phi_P|\eta_S\rangle\ ,\\[1ex]
\end{array}
\end{equation}
where
$|\eta_{NS}\rangle \equiv |u\bar u + d\bar d\rangle /\sqrt{2}$ and  
$|\eta_{S}\rangle \equiv |s\bar s\rangle$ 
are the non-strange and the strange basis states.
Most importantly, the two mixing angles now reduce to the one in Eq.~(\ref{mixingP}),
$\phi_P$, not in the good $SU(3)$ limit but in the much safer approximation of perfect
validity of the OZI-rule. 

These recent findings have been ignored in many previous treatments of $VP\gamma$ 
transitions and the related $\eta$-$\eta^\prime$  mixing pattern. 
For this reason and because of the rich set of data we now have at our disposal,
the purpose of this note is to present a detailed revision of those two issues. 

\section{A model for $VP\gamma$ $M1$ transitions}
\label{model}
We will work in a conventional quark model context and assume that pseudoscalar and 
vector mesons are simple quark-antiquark $S$-wave bound states.  
All these hadrons are thus extended objects with characteristic spatial extensions 
fixed by their respective quark-antiquark $P$ or $V$ wave functions.  
In the pseudoscalar nonet, $P=\pi, K, \eta, \eta^\prime$, 
the quark spins are antiparallel and the mixing pattern is given by Eq.~(\ref{mixingP}). 
In the vector case, $V=\rho, K^*, \omega, \phi$, the spins are parallel and mixing is 
similarly given by 
$|\omega\rangle \equiv \cos\phi_V |\omega_{NS}\rangle - \sin\phi_V |\omega_S\rangle$ and
$|\phi\rangle \equiv \sin\phi_V| \omega_{NS}\rangle + \cos\phi_V |\omega_S\rangle$,   
where $|\omega_{NS}\rangle$ and $|\omega_{S}\rangle$ are the analog non-strange and 
strange states, as before. 
We will work in the good $SU(2)$ limit with $m_u = m_d \equiv \bar{m}$ and with identical
spatial extension of wave functions within each $P$ and each $V$ isomultiplet.
$SU(3)$ will be broken in the usual manner taking constituent quark masses with
$m_s > \bar{m}$ but also, and this is a specific feature of our approach, 
allowing for different spatial extensions for each $P$ and $V$ isomultiplet.
Finally, we will consider that even if gluon annihilation channels may induce 
$\eta$-$\eta^\prime$ mixing, they play a negligible r\^ole in $VP\gamma$ transitions and
thus fully respect the usual OZI-rule. 

In our specific case of $VP\gamma$ $M1$ transitions, these generic statements translate 
into three characteristic ingredients of the model: 
\begin{itemize}
\item[{\it i)}] 
A $VP\gamma$ magnetic dipole transition proceeds via quark or antiquark spin-flip 
amplitudes proportional to $\mu_q=e_q/2m_q$. 
Apart from the  obvious quark charge values, this effective magnetic moment breaks
$SU(3)$ in a well defined way and distinguishes photon emission from strange or
non-strange quarks via $m_s > \bar{m}$.
\item[{\it ii)}] 
The spin-flip $V\leftrightarrow P$ conversion amplitude has then to be corrected by the 
relative overlap between the $P$ and $V$ wave functions. 
In older papers \cite{Isg,OD} a common, flavour-independent overlap was introduced.
Today, with a wider set of data, this new symmetry-breaking mechanism can be introduced 
without enlarging excessively the number of free parameters. 
\item[{\it iii)}] 
Indeed, the OZI-rule reduces considerably the possible transitions and their respective 
$VP$ wave-function overlaps: $Z_S$, $Z_{NS}$ and $Z_\pi$ characterize the 
$\langle\eta_S|\omega_{S}\rangle$, 
$\langle\eta_{NS}|\omega_{NS}\rangle =\langle\eta_{NS}|\rho\rangle$ and 
$\langle\pi|\omega_{NS}\rangle = \langle\pi|\rho\rangle$ spatial overlaps, respectively.
Notice that distinction is made between the $|\pi\rangle$ and $|\eta_{NS}\rangle$ 
spatial extension due to the gluon or $U(A)_1$ anomaly affecting the second state, 
but not between the  anomaly-free, non-strange vector states $|\rho\rangle$ and 
$|\omega_{NS}\rangle$.
Independently, we will also need $Z_K$ for the $\langle K|K^*\rangle$ overlap
between strange isodoublets. 
\end{itemize}

It is then a trivial task to write all the $VP\gamma$ couplings in terms of an
effective $g\equiv g_{\omega_{NS}\pi\gamma}$:
\begin{equation}
\label{couplings}
\begin{array}{c}
g_{\rho^0\pi^0\gamma}=g_{\rho^+\pi^+\gamma}=\frac{1}{3}\,g\ ,\\[2ex]
g_{\rho\eta\gamma}=g\,z_{NS}\,\cos\phi_P\ ,\\[2ex]
g_{\eta^\prime\rho\gamma}=g\,z_{NS}\,\sin\phi_P\ ,\\[2ex]
g_{\omega\pi\gamma}=g\,\cos\phi_V\ ,\\[2ex]
g_{\omega\eta\gamma}=\frac{1}{3}\,g\,
\left(z_{NS}\,\cos\phi_V\cos\phi_P -
      2\,\frac{\bar{m}}{m_s}\,z_S\,\sin\phi_V\sin\phi_P\right)\ ,\\[2ex]
g_{\eta^\prime\omega\gamma}=\frac{1}{3}\,g\,
\left(z_{NS}\,\cos\phi_V\sin\phi_P +
      2\,\frac{\bar{m}}{m_s}\,z_S\,\sin\phi_V\cos\phi_P\right)\ ,\\[2ex]
g_{\phi\pi\gamma}=g\,\sin\phi_V\ ,\\[2ex]
g_{\phi\eta\gamma}=\frac{1}{3}\,g\,
\left(z_{NS}\,\sin\phi_V\cos\phi_P +
      2\,\frac{\bar{m}}{m_s}\,z_S\,\cos\phi_V\sin\phi_P\right)\ ,\\[2ex]
g_{\phi\eta^\prime\gamma}=\frac{1}{3}\,g\,
\left(z_{NS}\,\sin\phi_V\sin\phi_P -
      2\,\frac{\bar{m}}{m_s}\,z_S\,\cos\phi_V\cos\phi_P\right)\ ,\\[2ex]
g_{K^{*0}K^0\gamma}=-\frac{1}{3}\,g\,z_K\,\left(1+\frac{\bar{m}}{m_s}\right)\ ,\\[2ex]
g_{K^{*+}K^+\gamma}=\frac{1}{3}\,g\,z_K\,\left(2-\frac{\bar{m}}{m_s}\right)\ ,\\[1ex]
\end{array}
\end{equation}
where we have redefined $z_{NS}\equiv Z_{NS}/Z_{\pi}$, $z_{S}\equiv Z_{S}/Z_{\pi}$ and
$z_{K}\equiv Z_{K}/Z_{\pi}$.
The normalization of the couplings is such that 
$g_{\omega\pi\gamma}=g\,\cos\phi_V =
2\,(\mu_u + \mu_{\bar{d}})\,Z_\pi\cos \phi_V =e\,Z_\pi\cos\phi_V/\bar{m}$ 
and the decay widths are given by 
\begin{equation}
\label{width}
\Gamma (V\rightarrow P\gamma)=
\frac{1}{3}\frac{g^2_{VP\gamma}}{4\pi}|{\bf p}_\gamma|^3=
\frac{1}{3}\Gamma (P\rightarrow V\gamma)\ , 
\end{equation}
where ${\bf p}_\gamma$ is the final photon momentum.

\section{Data fitting}
\label{datafitting}
The available experimental information on $\Gamma (V\rightarrow P\gamma)$ and 
$\Gamma (P\rightarrow V\gamma)$ partial widths is shown in the first column of 
Table \ref{table1} and has been taken exclusively from the recent 
PDG compilation \cite{PDG00}.
\begin{table}
\begin{center}
{\footnotesize
\begin{tabular}{|c|c|c|c|c|c|}
\hline
Transition & $\Gamma_{\rm exp}$(keV) & 
$\Gamma_{\rm fit1}$(keV) & $\Gamma_{\rm fit3}$(keV) &
$\Gamma_{\rm fit4}$(keV) & $\Gamma_{\rm fit5}$(keV)\\
\hline
$\rho^0\rightarrow\pi^0\gamma$       & $102\pm 26$ & $75\pm 4$
                                     & $67\pm 4$ & $74\pm 4$ & $71\pm 9$\\
$\rho^+\rightarrow\pi^+\gamma$       & $68\pm 7$ & $74\pm 4$
                                     & $67\pm 4$ & $74\pm 4$ & $71\pm 8$\\
$\rho^0\rightarrow\eta\gamma$        & $36^{+12}_{-14}$ & $46\pm 6$
                                     & $53\pm 3$ & $49\pm 5$ & $44\pm 6$\\
$\eta^\prime\rightarrow\rho^0\gamma$ & $60\pm 5$ & $59\pm 10$
                                     & $58\pm 5$ & $58\pm 13$ & $58\pm 9$\\
$\omega\rightarrow\pi^0\gamma$       & $717\pm 43$ & $708\pm 36$
                                     & $637\pm 23$ & $704\pm 35$ & $720\pm 42$\\
$\omega\rightarrow\eta\gamma$        & $5.5\pm 0.8$ & $5.1\pm 0.8$
                                     & $5.9\pm 0.4$ & $5.5\pm 0.6$ & $5.2\pm 0.5$\\
$\eta^\prime\rightarrow\omega\gamma$ & $6.1\pm 0.8$ & $6.7\pm 0.8$
                                     & $6.9\pm 0.6$ & $6.4\pm 1.3$ & $6.9\pm 0.8$\\
$\phi\rightarrow\pi^0\gamma$         & $5.6\pm 0.5$ & $5.6\pm 0.6$
                                     & $5.6\pm 0.5$ & $5.6\pm 0.6$ & $5.6\pm 0.8$\\
$\phi\rightarrow\eta\gamma$          & $57.8\pm 1.5$ & $58\pm 13$
                                     & $58.0\pm 6.7$ & $58\pm 11$ & $57.7\pm 6.9$\\
$\phi\rightarrow\eta^\prime\gamma$   & $0.30^{+0.16}_{-0.14}$ & $0.37\pm 0.08$
                                     & $0.43\pm 0.04$ & $0.25\pm 0.06$ & $0.36\pm 0.03$\\
$K^{*0}\rightarrow K^0\gamma$        & $116\pm 10$ & $117\pm 13$
                                     & $124\pm 6$ & $115\pm 11$ & $$\\
$K^{*+}\rightarrow K^+\gamma$        & $50\pm 5$ & $50\pm 6$
                                     & $56\pm 4$ & $50\pm 5$ & $$\\
\hline
\end{tabular}
}
\end{center}
\caption{Comparison between the experimental values $\Gamma_{\rm exp}$ 
for the various $VP\gamma$ transitions taken from Ref.~\protect\cite{PDG00} 
and the corresponding predictions
$\Gamma_{\rm fit1}$, $\Gamma_{\rm fit3}$, $\Gamma_{\rm fit4}$ and
$\Gamma_{\rm fit5}$ from Eqs.~(\protect\ref{fit1}), (\protect\ref{fit3}),
(\protect\ref{fit4}) and (\protect\ref{fit5}), respectively.}
\label{table1}
\end{table}
A fit to these data with the couplings of our model, Eq.~(\ref{couplings}), leads to
the predictions $\Gamma_{\rm fit1}$ listed in the second column of Table \ref{table1}. 
The values of our seven free parameters are found to be 
\begin{equation}
\label{fit1}
\begin{array}{c}
g=0.70\pm 0.02\ \mbox{GeV$^{-1}$}\ ,\ \ \ m_s/\bar{m}=1.24\pm 0.07\ ,\\[2ex]
\phi_P=(37.7\pm 2.4)^\circ\ ,\ \ \ \phi_V=(3.4\pm 0.2)^\circ\ ,\\[2ex]
z_{NS}=0.91\pm 0.05\ ,\ \ \ z_S=0.89\pm 0.07\ ,\ \ \ z_K=0.91\pm 0.04\ .\\[1ex]
\end{array}
\end{equation}
The quality of the fit is excellent, $\chi^2/d.o.f.=3.2/5\simeq 0.6$.
The fitted values for the two mixing angles $\phi_P$ and $\phi_V$ are in good agreement
with most results coming from other analyses using complementary information.
Our values for $g$ and $m_s/\bar{m}$ are also quite reasonable although their comparison
with those from alternative studies is much more model dependent \cite{Del}.
The three $z$'s are specific of our approach and different from unity,
the approximate value assumed in previous analyses.
To further investigate this last issue, we have performed a second fit to the same
twelve data fixing now $z_{NS}=1$ and $z_{S}=z_{K}^2$.
The quality of the fit, $\chi^2/d.o.f.=6.8/7\simeq 1$, is substantially reduced.
This shows that allowing for different overlaps of quark-antiquark wave functions and,
in particular, for those coming from the gluon anomaly affecting only the $\eta$ and 
$\eta^\prime$ singlet component, has indeed some relevance. 

As previously stated, a few of the twelve experimental data we are dealing with come
from difficult Primakoff-effect analyses and could be affected by large uncertainties.
The neutral and charged $K^*\rightarrow K\gamma$ transitions, for instance, 
have been measured only by one and two experimental groups respectively,
and seem to need further confirmations. 
For these reasons, and also to allow later for easier comparison with work by other 
authors, we have performed a new fit ignoring the two $K^*\rightarrow K\gamma$ transition
information.
This new fit requires
\begin{equation}
\label{fit2}
\begin{array}{c}
g=0.70\pm 0.02\ \mbox{GeV$^{-1}$}\ ,\\[2ex]
\phi_P=(37.7\pm 2.4)^\circ\ ,\ \ \ \phi_V=(3.4\pm 0.2)^\circ\ ,\\[2ex]
z_{NS}=0.91\pm 0.05\ ,\\[1ex]
\end{array}
\end{equation}
whereas $m_s/\bar{m}$ and $z_S$ always appear in the combination 
$z_S\,\bar{m}/m_s$ fitted to $0.72\pm 0.04$. 
The quality of the fit, $\chi^2/d.o.f.=3.2/5\simeq 0.6$,
is as good as in the previous global fit and the results are practically identical.
The adequacy of our treatment and the values of its main parameters are therefore 
insensitive to eventual modifications of future and desirable new data on 
$K^* \rightarrow K \gamma$ transitions.

From Eq.~(\ref{couplings}) one can immediately deduce the ratios  
\begin{equation}
\label{four}
\begin{array}{c}
\frac{g_{\rho\eta\gamma}}{g_{\eta^\prime\rho\gamma}}=\cot\phi_P\ ,\\[2ex]
\frac{g_{\omega\eta\gamma}}{g_{\eta^\prime\omega\gamma}}\simeq 
\cot\phi_P\,\left(1-4\,\frac{\bar{m}}{m_s}\,\tan\phi_V\right)\ ,\\[2ex]
\frac{g_{\phi\eta\gamma}}{g_{\phi\eta^\prime\gamma}}\simeq 
-\cot\phi_P\,\left(1-4\,\frac{m_s}{\bar{m}}\,\tan\phi_V\right)\ ,\\[2ex]
\frac{g_{K^{*0}K^0\gamma}}{g_{K^{*+}K^+\gamma}}=
\frac{1+m_s/\bar{m}}{1-2m_s/\bar{m}}\ ,\\[1ex]
\end{array}
\end{equation}
where the two approximate expressions are remarkably accurate as a consequence of the
results of our fits: $\tan \phi_V\ll 1, \sin 2\phi_P\simeq 1, z_{NS}\simeq z_{S}$. 
The value of $m_s/\bar{m}$ depends mainly on the fourth ratio in Eq.~(\ref{four}) 
involving only $K^*$--$K$ transitions and its $z_K$ independence was 
appreciated years ago by Sucipto and Thews \cite{ST}.
The first three ratios are essential to fix the pseudoscalar nonet mixing angle $\phi_P$.
Again, they are practically $z$-independent, whereas they turn out to depend 
significantly on the ratio $f_\eta/f_{\eta^\prime}$ in alternative approaches.
This feature could have some relevance when extracting the value for $\phi_P$ and 
comparing with results from other authors, as we now proceed to discuss.

\section{Comparison with other approaches} 
\label{comparison}
As just stated, several recent analyses of $VP\gamma$ transitions \cite{BGP,Has}
introduce symmetry-breaking terms in such a way that the various $VP\gamma$ amplitudes
turn out to be dependent on the corresponding $P$ decay constant $f_P$.
The clear advantage of this procedure is that valuable information on 
$P\rightarrow\gamma\gamma$ transitions can be treated within the same context.
A serious drawback, as already mentioned in the Introduction, is that the complicated
two-angle dependence of $f_{\eta, \eta^\prime}$ on $f_{8, 0}$ is hard to take into account
and usually ignored 
(an exception is the recent treatment of $\phi\rightarrow\eta\gamma, \eta^\prime\gamma$ 
transitions via QCD sum rules in Ref.~\cite{FP}). 
A drastic solution for this problem could consist in fixing all $f_P$'s to the same 
unbroken value, thus accepting that one has no control on this part of the 
symmetry-breaking mechanism and that the final results are just a rough estimate.
This is equivalent to the treatment in Refs.~\cite{BS,BES} or to the present one fixing 
all the $z$'s to unity.
In this case, a fit to the twelve experimental entries of Table \ref{table1} leads 
to the estimates $\Gamma_{\rm fit3}$ listed in its third column.
The quality of the fit now decreases to $\chi^2/d.o.f.= 10.4/8\simeq 1.3$
but the  values of the main parameters are quite consistent with our previous ones:  
\begin{equation}
\label{fit3}
\phi_P=(35.6\pm 1.8)^\circ\ ,\ \ \ m_s/\bar{m}=1.27\pm 0.05\ .
\end{equation} 
 
Other authors \cite{Ben,GMR} have proposed different symmetry-breaking mechanisms
inspired in earlier work by O'Donnell \cite{OD}. 
Rather than assuming the nonet symmetry ordinarily associated to quark model ideas, 
Benayoun {\it et al.}~\cite{Ben} include a nonet symmetry breaking parameter, $x$, 
in their approach. 
The expressions for their coupling constants follow from those in Eq.~(\ref{couplings})
once we put $z_{NS}=z_{S}=1$ and substitute $\cos\phi_P$ and $-\sin\phi_P$ in the 
couplings involving an $\eta$ meson by $X^{NS}_\eta$ and $X^{S}_\eta$, respectively,
with $X^{NS}_\eta=\cos\phi_P\,(1+2x+\sqrt{2}(1-x)\tan\phi_P)/3$, 
$X^{S}_\eta=-\sin\phi_P\,(2+x+\sqrt{2}(1-x)\cot\phi_P)/3$;
the couplings involving an $\eta^\prime$ meson can then be obtained from the latter 
substituting $\cos\phi_P$ and $-\sin\phi_P$ by $\sin\phi_P$ and $\cos\phi_P$,
respectively, as required by  Eq.~(\ref{mixingP}).
The four ratios (\ref{four}) follow then at leading order of symmetry-breaking.
Also, a global fit leads again to the excellent results $\Gamma_{\rm fit4}$ 
listed in the fourth column of Table \ref{table1} ($\chi^2/d.o.f.=3.1/6\simeq 0.5$) 
and to the values
\begin{equation}
\label{fit4}
\phi_P=(40.0\pm 2.8)^\circ\ ,\ \ \ m_s/\bar{m}=1.25\pm 0.06\ .
\end{equation}

Finally, we discuss another source of $SU(3)$-breaking corrections suggested in other
recent treatments by Fr\`ere {\it et al.}~\cite{BFT,EF} 
and by Feldmann {\it et al.}~\cite{FKS,Fel}.
It consists in the introduction of different annihilation constants, $f_V$, for the 
various vector mesons, quite in line with the different $f_P$'s simultaneously used in 
the pseudoscalar sector, both accounting for the values of the respective wave functions
at the origin.
In a sense, the  $f_V f_P$ factor appearing in the corresponding $VP\gamma$ transition
(see Refs.~\cite{FKS,Fel} for details) are then related to our $Z$ factor accounting 
similarly for the wave-function overlap. 
The independent symmetry-breaking factor $m_s/\bar{m}$
---which is essential to adjust the ratio between the two $K^*$--$K$ transitions---
is not contemplated in Refs.~\cite{FKS,Fel}, thus precluding the immediate possibility 
of an acceptable global fit.
The other first three ratios in Eq.~(\ref{four}) are easily reproduced and an excellent 
fit is obtained if the two kaonic channels are excluded, as shown in the final column of
Table \ref{table1}. 
The quality of the fit is $\chi^2/d.o.f.=2.9/3\simeq 1$ and the mixing angle is
\begin{equation}
\label{fit5}
\phi_P=(38.1\pm 2.5)^\circ\ .
\end{equation}

\section{Conclusions}
\label{conclusions}
The old and widespread belief that simple quark-model ideas are quite appropriate to 
describe $VP\gamma$ transitions has been confirmed by the present analysis.  
Indeed, the rather solid set of data now available covering {\it all} possible 
$VP\gamma$ transitions between the pseudoscalar and vector meson nonets has been shown 
to be easily described in terms of the basic model implemented with various 
symmetry-breaking mechanisms.
Distinction among the latter seems feasible by comparing future and more accurate
measurements of $\Gamma(\phi\rightarrow\eta^\prime\gamma)$ with the various predictions
shown in Table \ref{table1}.
However, quite independently of the details of these mechanisms one can safely conclude 
from Eqs.~(\ref{fit1}), (\ref{fit2}), (\ref{fit4}) and (\ref{fit5}) that the value of 
the $\eta$-$\eta^\prime$ mixing angle $\phi_P$ deduced from $VP\gamma$ data has to be
in the range $37.5^\circ$--$39.5^\circ$. 

More specifically, we propose the value $\phi_P=(37.7\pm 2.4)^\circ$ following from our
own treatment of $SU(3)$-breaking effects.
This treatment, in line with the recent approach by 
Feldmann {\it et al.}~\cite{FKS,Fel} emphasizing the r\^ole played by the non-strange 
and strange components of the $\eta$ and $\eta^\prime$ mesons, circumvents the 
difficulties encountered in other $\eta$-$\eta^\prime$ mixing analyses.
Moreover, $SU(3)$-breaking effects originated by the flavour-dependence in the various
$VP$ wave-function overlaps are taken into account.  
This flavour-dependence turns out to be relevant and contains useful information on
the spatial extension of the $P$ and $V$ mesons. 

\section*{Acknowledgements}
Work partly supported by the EEC, TMR-CT98-0169, EURODAPHNE network.
M.~D.~S.~is grateful for the hospitality received at IFAE 
where part of this work was done.

\end{document}